\documentstyle[psfig]{laa}   

\begin{document}
\thesaurus{06(08.09.2 CM 
Dra,08.16.2,08.02.2,08.12.2,08.06.1,03.20.4)} 
\title{Near-Term Detectability of Terrestrial Extrasolar Planets: 
TEP Network Observations of CM Draconis}

\author{H.J. Deeg\inst{1}, L.R. Doyle\inst{2}, V.P. 
Kozhevnikov\inst{3}, E.L. Mart\'\i n\inst{1,}\thanks{{\it present 
address:\/}\ University of California at Berkeley, 601 Campbell Hall, 
CA 94720}, B. Oetiker\inst{4}, E. Palaiologou\inst{5}, J. 
Schneider\inst{6}, C. Afonso\inst{6}, E.W. Dunham\inst{7}, J.M. 
Jenkins\inst{2}, Z. Ninkov\inst{8}, R.P.S. Stone\inst{9}, P.E. 
Zakharova\inst{3}}

\institute{Instituto de Astrof\'{\i}sica de Canarias, E-38200 La 
Laguna, 
Tenerife, Spain, hdeeg@bigfoot.com
\and SETI Institute, MS 245-3, NASA Ames Research Center, Moffett 
Field, CA 94035, USA, doyle@gal.arc.nasa.gov
\and Astronomical Observatory, Ural State University, Lenin Ave. 51, 
Ekaterinburg, 620083, Russia
\and University of New Mexico, Dept. of Physics and Astronomy, 
Albuquerque, NM 87131, USA
\and University of Crete, Skinakas Observatory, P.O. Box 1527, 
Heraklion 71110, Crete, Greece
\and CNRS-Observatoire de Paris, 92195 Meudon, France
\and Lowell Observatory, Flagstaff, AZ 86001, USA
\and Center for Imaging Science, Rochester Institute of Technology, 
Rochester, NY 14623-5604, USA
\and University of California, Lick Observatory, Mount Hamilton, CA 95040, USA}

\offprints{H.J. Deeg}

\date{Received 17 October 1997; accepted 9 June 1998}
\maketitle
\begin{abstract}

Results from a photometric search for extrasolar planetary transits
across the eclipsing binary CM Dra are presented.  The TEP (Transits
of Extrasolar Planets) network has observed this star since 1994, and
a lightcurve with 617 hours of coverage has been obtained. The data
give a complete phase coverage of the CM Dra system at each of the 3
years of observations, with a noise of less than 5 mmag. New epoch and
period values for CM Dra are derived, and a low flare rate of 0.025
$hr^{-1}$ has been confirmed. The absence of periodic variations in
eclipse minimum times excludes the presence of very massive planets
with periods of less than a few years. The lightcurve was visually
scanned for the presence of unusual events which may be indicative of
transits of extrasolar planets with 'massive earth' sizes. Six
suspicious events were found which are being followed up for future
transits, by planets with sizes between 1.5 and 2.5 $R_{E}$ (Earth
Radii).  However, none of these events has amplitudes compatible with
planets larger than 2.5 $R_{E}$.  Coplanar planets larger than 2.5
$R_{E}$ and with orbital periods of less than 60 days can therefore be
ruled out with a confidence of about 80\%.  Planets smaller than 1.5
$R_{E}$ cannot be detected in the data without a sub-noise detection
algorithm. A preliminary signal detection analysis shows that there is
a 50\% detection confidence for 2 $R_{E}$ planets with a period from
10 to 30 days with the current data.  This data-set demonstrates that
it is possible to detect terrestrial sized planets with ground based
photometry, and that strong constraints on the sizes of planets
orbiting in the plane of the CM Dra system can be set.

\end{abstract}

\begin{figure*}
%below, for preprint (2 col) style
\centerline{\psfig{figure=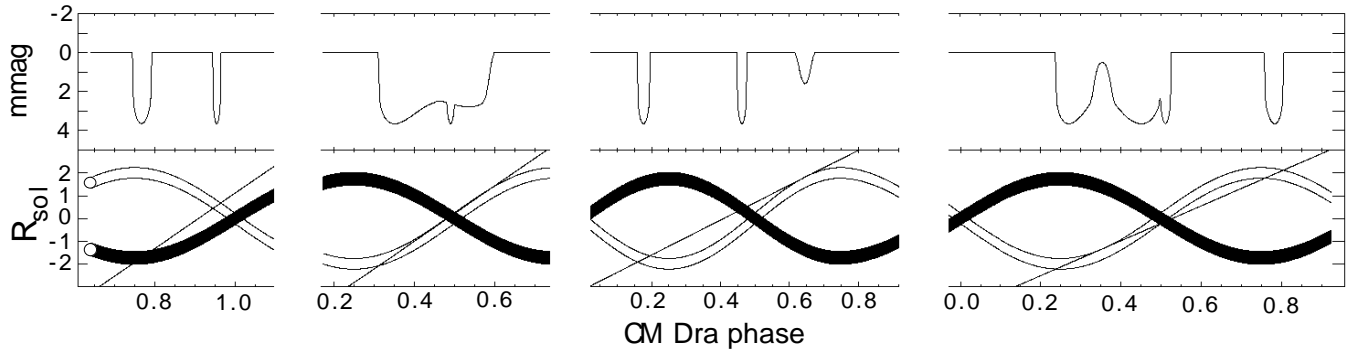,width=18cm}}
%below, for referee (1 col) style
%\centerline{\psfig{figure=H0739F1.eps,width=15cm}}

\caption{\label{fig:transshapes}Model lightcurves from planetary
transits across CM Dra. The upper graph gives the brightness of the CM
Dra system, normalized to an off-transit magnitude of zero. A model
planet with 2 R$_E$ causes transits with a maximum brightness loss of
3.8 mmag (0.35 \%). Mutual binary eclipses are removed here. The lower
graphs show the elongation (in Solar radii) of the two CM Dra
components (CM Dra A: black band, B: white band) and the model-planet
(thin line) from the common barycenter. These graphs are of the same
style as the well-known diagrams of the positions of Jupiter's moons,
and the thickness of the bands is scaled to the sizes of CM Dra's
components.  From the left to the right, the two leftmost panels show
transits caused by a planet with an orbital period of 9 days. The
leftmost panel shows the normal case, with two short transits
separated by several hours (0.1 CM Dra phase unit corresponds to
$\approx$ 3 hours). The second panel shows the rarer case, where the
planet transits when CM Dra is close to a mutual eclipse (here shown a
secondary one at phase 0.5), and long transits occur. The complicated
shape of this transit results from the planet covering zones of
different surface brightness on CM Dra (considering limb-darkening and
the planet's ingress/egress). The two rightmost panels show transits
from a planet with a 36 day period. Such a planet has a transversal
velocity that is slower than CM Dra's components, and multiple
transits with complicated shapes are more likely to occur.}
\end{figure*}

\section{Introduction}

In this paper we describe the observations and analysis that have been
performed in a search for photometrically detectable signals from the
presence of extrasolar planets around the eclipsing binary CM
Draconis. This is the first long-term observational application of the
transit method for the detection of extrasolar planets. The transit
method is based on observing small drops in the brightness of a
stellar system, resulting from the transit of a planet across the disk
of its central star.  Such transits would cause characteristic changes
in the central star's brightness and, to a lesser extend, color.  The
depth of a transit is proportional to the surface area of the planet,
and the duration of a transit is indicative of the planet's
velocity. If the central star's mass is known, the distance and period
of the planet can then be derived.  Once repeated transits of the same
planet are observed, the period can be obtained with great precision.
The transit method was first proposed by Struve (1952); later
developments are described in \cite{rose71,boru84,deeg97}. Previous
observational tests have been prevented by the required photometric
precision (which is about 1 part in 10$^{5}$ in the case of an
Earth-sized planet transiting a sun-like star), and by the generally
low probability that a planetary plane is aligned correctly to produce
transits. This probability of orbital alignment is about 1\% for
planetary systems similar to our solar system.  An observationally
appealing application is available with close binary systems, where
the probability is high that the planetary orbital plane is coplanar
with the binary orbital plane, and thus in the line sight. This makes
the observational detection of planetary transits feasible in systems
with an inclination very close to 90$^{o}$ (\cite{schn90,schn95}).
Furthermore, repeated planetary transits across the binary's
components will result in unique sequences of transit-lightcurves,
whose exact shape depends on the phase of the binary system at the
time of the planetary transit (Fig.~\ref{fig:transshapes}). Jenkins et
al. (1996) demonstrated, that these unique sequences can be used for
the detection of planetary signatures with amplitudes below the noise
of the observed lightcurves, if signal detection techniques based on
cross correlations with model lightcurves are applied.

The near ideal characteristics of the eclipsing binary system CM Dra
for an observational test on the presence of planets has been
suggested by Schneider and Doyle (1995). The CM Dra system is the
eclipsing binary system with the lowest mass known, with components of
spectral class dM4.5/dM4.5 (see \cite{lacy77}, for all system
elements). The total surface area of the systems' components is about
12\% of the sun's, and the transits of a planet with 3.2 R$_E$,
corresponding to 2.5\% of the volume of Jupiter, would cause a
brightness drop of about 0.01 mag, which is within easy reach of
current differential photometric techniques. The low temperature of CM
Dra also implies that planets in the thermal regime of solar system
terrestrial planets would circle the central binary with orbital
periods on the order of weeks. This allows for a high detection
probability of planetary transits by observational campaigns with
coverages lasting more than one planetary period. Planets with orbital
periods of 10 - 30 days around CM Dra are especially interesting,
since they would would lie within the habitable zone, which is the
region around a star where planetary surface temperatures can support
liquid water, and therefore the development of organic life (see
Kasting et al. 1993; Doyle 1996).  CM Dra is relatively close (17.6
pc) and has a near edge-on inclination of 89.82\degr. With this
inclination, coplanar planets within a distance of CM Dra of $\approx$
0.35 AU will cause a transit event. This maximum distance corresponds
to a circular orbit with a period of about 125 days. There is also a
low probability of observing orbits from planets inclined out of CM
Dra's binary orbital plane, if the ascending or descending nodes of
the planetary orbits are precessing across the line of sight
(\cite{schn94}). The observations of CM Dra presented in this paper
are therefore the first attempt to obtain observational evidence of
the existence of sub-Jupiter sized planets around main-sequence binary
stars, and to evaluate the probability that such detections are
possible. For these observations we used differential CCD-photometry
and employed 1m class telescopes. To obtain sufficient observational
coverage, the 'TEP' (Transits of Extrasolar Planets) network was
formed with the participation of several observatories in
1994. Preliminary accounts of TEP network observations are given by
Doyle et al. (1996), Martin et al. (1997) and Deeg et al. (1997). A
list of the observatories that have been participating is given in
Table~\ref{tab:scopes}.

\begin{table*}
\caption{\label{tab:scopes} TEP Network Telescopes and their Location}
\begin{center}
\begin{tabular}{lccccc} 
\hline
Telescope&Location&f ratio&Longitude&Latitude&Altitude\\
\hline
IAC 80cm&Iza\~{n}a, Tenerife&f/14.4&16\degr31'W&28\degr18'N&2385m\\
 &Canary Islands\\
JKT 1m&Roque Muchachos&f/15&17\degr 53'W&28\degr 46'N&2365m\\
&La Palma, Canary Isl.\\
INT 2.5m&Roque Muchachos&f/3.29&17\degr 53'W&28\degr 46'N&2340m\\
&La Palma, Canary Isl.\\
Mees 24''&Rochester, NY&f/15.2&77\degr 25' W&42\degr 40' N&690m\\
Capilla 24''&Albuquerque, NM&f/15.2&106\degr 24'W&34\degr 42'N&2835m\\
Crossley 36''&Lick Observatory, CA&f/5.8&121\degr 39W&37\degr 
20'N&1210m\\
Kourovka 0.7m&Ural University&f/14.3&59\degr 30'E&57\degr  03'N&320m\\
&Ekaterinburg, Russia\\
WISE 40''&Negev, Israel&f/7&34\degr 45' E&30\degr 36' N&875m\\
Skinakas 1.3m&Skinakas Obsv., Crete&f/8&24\degr 53' E& 35\degr 23' 
N&1752m\\
OHP 1.2m&Obsv. Haute Provence&f/6&5\degr 42' E&44\degr N&650m\\
&France\\
\hline
\end{tabular}
\end{center}
\end{table*}

\begin{table*}
\caption{\label{tab:CCDs} CCD systems}
\begin{center}
%\tiny
\begin{tabular}{llcccccccc}
\hline
Telescope&CCD system&Pixel&field of&$t_{exp}^{\ 
1}$&Duty$^2$&Read-noise\\
&&size&view&(sec)&fraction&(electrons)\\
&&(arcsec)&(arcmin)&&\\
\hline
IAC 80cm$^3$&Tek 1024(1994)&0.43&7.2'&60&0.50&5.8\\
&''\ \  (1995,96)&''&''&180&0.84&5.8\\
JKT 1m&Tek 1024$^4$(\# 4)&0.33&5.65&50&0.20&6.9\\
INT 2.5m&Tek 1024$^4$ (\# 3)&0.59&10.0&80&0.59&6.1\\
Mees 24''&Kodak KAF4200$^5$&0.40$^{6}$&6.82&300&0.91&17.1\\
Capilla 24''&RCA SID50EX&0.67&3.57x5.72&120&0.77&57\\
Crossley 36''$^3$&custom$^{7}$ CCD(1994)&0.58&20&90&0.58&19\\
&custom$^{7}$ CCD(1995)&''&''&60&1.0$^7$&19\\
&KAF 4200(1996)&0.71$^{6}$&12.1&120&0.76&17.1\\
Kourovka 0.7m&two-star&n/a&20 $\oslash$&128&1.0&n/a\\
&photometer\\
WISE 40''&Tek 1024&0.7&11.9&240&0.86&6\\
Skinakas1.3m&TH 7896A&0.398&6.79&60&0.60&6\\
OHP 1.2m&TEK512CB&0.78&6.6&48&0.60&6.8\\
\hline
\end{tabular}
\end{center}
$^1$$t_{exp}$ is the typical exposure time used for a  CCD image of 
CM Dra\\
$^2$The Duty fraction is the fraction of time the camera is collecting 
light while observing the object. It is given by: 
$t_{exp}/(t_{exp}+t_{read}+t_{disk})$, where $t_{read}$ is the CCD 
read-out time, and $t_{disk}$ is the time to save an image to the 
disk.\\
$^3$Different cameras or settings were used at these telescopes 
through the years\\
$^4$A nonlinearity not exceeding 140 e- was corrected from the CCD's 
linearity calibration curve.\\
$^5$ Nonlinearity was corrected using precints by Deeg and Ninkov 
(1996)\\
$^6$ Used CCD camera in 2x2 bin mode\\
$^7$ This CCD system is described in \cite{dunham95}.  Duty fraction 
of 1.0 results from use of CCD frame-transfer mode\\

\end{table*}

\section{Observations}
 
The vast majority of the observations has been performed with 
CCD-equipped telescopes with sizes ranging from 0.6m to 1.2m.  
Although the techniques for high precision photometric work using 
photo multipliers have become very refined (e.g. \cite{youn91}), 
such work has concentrated mainly on objects significantly brighter than 
CM Dra, which is magnitude 11.07 in R-band.  With one - or two - channel 
photometers requiring frequent sky or standard-star observations, their 
duty cycles are relatively low.  The use of CCD cameras allows the 
simultaneous observation of several reference stars in the same field 
as CM Dra, and the duty cycle is only limited by the time to read out 
the CCD and to save the image to disk, both together being on the 
order of one minute.  This allows tracing the lightcurve with 
measurements spaced 2-4 minutes apart.  The close spacing of 
measurements is important in order to recognize observed brightness 
variations 
as potential planetary transits.  Kjeldsen \& Frandsen (1992) have 
demonstrated the feasibility of the use of CCD's in high-precision 
time-resolved photometry, and emphasized their usefulness for the 
study 
of low-amplitude variables (see also \cite{gill88}).  In addition, 
occasional small-scale variations in atmospheric transmission properties 
(for example, atmospheric density waves) that could appear as 
planetary transit events in conventional photometry can be isolated 
with a wide-field CCD photometric system, using a significant number 
of comparison stars within the field.  We would also like to emphasize 
CCD's relative ease of use, as the simultaneous observations of 
reference stars allows one to leave the telescope pointed towards CM 
Dra throughout a whole observing night, which is a tremendous 
advantage on simpler telescopes without computer control.

\begin{figure*}
\centerline{\psfig{figure=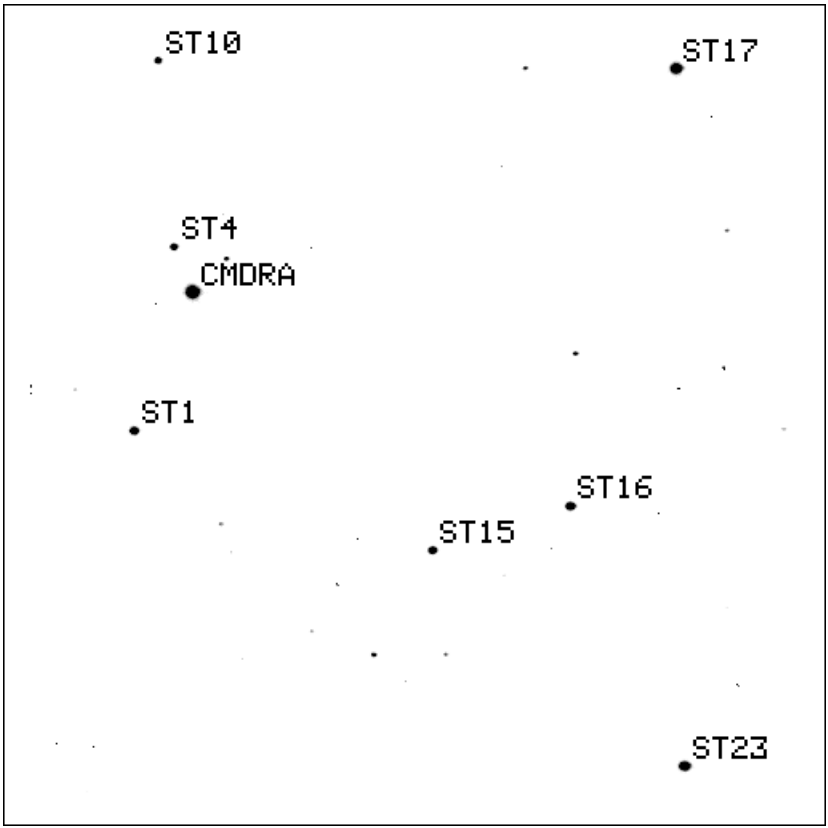,bbllx=0pt,bblly=0pt,bburx=614pt,bbury=614pt,width=13cm,height=13cm,rwidth=14cm,rheight=14cm}}

\caption{\label{fig:ccdfield}
Field of CM Dra with the most commonly used reference stars marked. N 
is up and E is left. The side length of the field is 7.2 arcmin}
\end{figure*}

A list of the properties of the CCD systems used at the various
telescopes is given in Table~\ref{tab:CCDs}. Only at Kourovka
Observatory a two-star photometer was employed. In all cases - except
as noted - observations were taken through a standard R filter. The
CCDs had a field of view of at least 5' side-length, which allowed us
to observe within the images of CM Dra at least 5 reference stars
simultaneously - these were normally the stars numbered 1, 4, 15, 16
and 17 in Fig.~\ref{fig:ccdfield}. Exposure times at the various
telescopes ranged from 30 to 300 sec, depending on the telescope's
size and the dynamic range of the CCD (see Table~\ref{tab:CCDs}). A
maximum exposure time of 300 sec was set to acquire lightcurves with
sufficient temporal resolution. At the two-star photometer at Kourovka
Observatory, star HD 150172 was observed simultaneously as a
reference.  This star is very similar in brightness to CM Dra, and
photometric counts were recorded every 128 seconds.

\section{Reduction}

The CCD images were bias subtracted and flatfielded using the common
procedures in the IRAF software.  As the object field is uncrowded,
aperture photometry was found to deliver more consistent results than
methods based on point-spread-function (psf) fitting. Depending on the
different telescopes' fields of view, between 5 and 9 suitable field
stars were used as reference "standard" stars.  To perform aperture
photometry with a maximum signal-to-noise, optimum aperture sizes
(Howell, 1989) for each star were determined.  Frame-to-frame
variations in the size of the stars' psf result from changes in seeing
conditions or from changes in the telescope's focusing throughout a
night. To correct for this on each frame, the psf of CM Dra (which was
the brightest star in the field) was fitted by a circular
Gaussian. All apertures are then expressed in multiples of the FWHM of
this psf. These multiples were kept constant throughout a night's
data. A suite of IRAF tasks for time-series photometry with optimized
apertures in uncrowded fields, named '\emph{vaphot}', was developed
for these reductions. \emph{vaphot} is available upon request from
H. Deeg.

The reference magnitude was based on the sum of the flux of the
reference stars, against which the differential magnitude of CM Dra
was calculated.  Individual reference stars may have unusual
brightness variations in some nights, which may result from intrinsic
variability or from flatfielding residuals, as described later.  To recognize these
variations, the difference between each reference star's magnitude and
the summed reference magnitude was checked for variability, and often
the rejection of one or two reference stars led to improved light
curves of CM Dra with lower noise (see Deeg et al., 1997, for an
example).  The resulting lightcurves were cleaned of obviously
erroneous measurements, as well as of events which are most likely
flares of CM Dra (see Sect. 4.3).  The differential magnitude was
then scaled such that CM Dra's average magnitude outside of mutual
binary eclipses was zero.  All the steps described in this paragraph
were performed independently for each night's observations.

The large color differential between the M4 stars composing CM Dra (V-R = 1.8) and the
reference stars (V-R = 0.55 to 0.7, except reference star 4: V-R =
0.33, which is a white-dwarf proper-motion companion of CM Dra at a
distance of $d \sin i =445 AU$; \cite{lacy77}) caused slow
airmass-related changes in CM Dra's brightness from differential
extinction.  In lightcurves with apparent slow variations caused by
differential extinction, these variations were removed by subtraction
of a fit, which was either a linear or a 2nd order polynomial fit to
the off-eclipse lightcurve. With some rare exceptions, the events
caused by a possible planetary transit occur on the time-scale of an
hour, with ingress/egress lasting on the order of 10 minutes. The
removal of the extinction slopes has therefore only a small effect on
the signal content of the lightcurve resulting from planetary
transits. This removal will also suppress amplitude variations from
star-spots, which occur on approximately the same time-scale as the
extinction, since the period of CM Dra's components is very likely
locked to the binary period of 1.27 days. Final lightcurves were
produced in 3 versions: A 'raw' one containing spurious points and
flares, a 'cleaned' one where these events have been removed, and a 'fitted' one
where the nightly slopes have been removed, as described above.

At some observatories the reduction procedure to obtain the 'raw'
lightcurve had to be modified: The data from the CCD at the Rochester
telescope exhibited a nonlinearity (\cite{deeg95}), which required a
correction step before flat fielding. The raw reduction of the data
from the Skinakas telescope was performed independently, using the 
MIDAS
software (Palaiologou, personal communication). The photometer data
from Kourovka observatory only required subtraction of the reference
star's magnitude, followed by removal of the nightly extinction slope.

%24 JAN: BELOW, ADDITIONAL DISCUSSION OF NOISE
Of special interest for the detection of planetary transits are noise
and error sources which can cause deviations in the data that may
appear similar to planetary transits.  The two major sources for errors with time
scales of transit events (~$\stackrel{>}{_\sim}$~40~min) are (i)
atmospheric instabilites and (ii) flatfielding errors.

(i) Atmospheric effects: 
The reference stars where traced for changes in their relative 
brightness amongst each other.  If such changes occured above the 
normal noise, these particular data were rejected.  As mentioned, CM 
Dra is a much redder star than any of the reference stars in the 
field, whose colors are all within a relatively narrow range.  The brightness ratio between CM Dra and the reference 
stars is therefore more sensitive to second order extinction changes 
(i.e.  changes in the color-dependency of the extinction within the 
bandpass of the R-filter) as are the brightness ratios among the 
reference stars themselves.  In rare cases, it may be feasible, that 
temporary effects (for example, a band of very fine cirrus, or dust) 
in the atmospere can cause strong second order extinction changes, 
without sufficient first-order extinction variations to warrant a 
rejection of the data.

(ii) Flatfielding: Errors from flatfielding may appear if the spatial 
distribution of the stellar light on the CCD undergoes positional 
changes between images.  This happens if the telescope tracking allows 
the stars to appear in slightly different positions in subsequent 
exposures.  Imperfect flatfielding corrections will then appear as 
brightness variations among the stars.  Flatfielding effects may 
strongly affect the data, if a star happened to move over a bad CCD 
pixel or column, or over a dust-corn on an optical surface.  Although 
care was taken to keep these `features' away from any star, and 
especially from CM Dra, sometimes they may not have been recognized, 
or dust-corns may have changed their positions.  If such a 
flatfielding variation affected any one of the reference stars, it 
caused an unusual change in its brightness relative to the other 
reference stars, and this reference star was not used in that night's 
summed reference magnitude.  More difficult are cases in which the 
program star (CM Dra) may have been affected by flatfielding 
variations.  To avoid this, we tracked the positional changes 
throughout each night, and rejected data with suspicious brightness 
variations of CM Dra, \emph{if} they correlated with positional 
movements.

Even in a well-tracking telescope, where the center positions of the 
stars do not notably change throughout a night, flatfielding effects 
may appear if changes in the seeing cause variations in the relative 
illumination among the CCD pixels.  If seeing was monotonically 
de- or increasing throughout a night (which was the most frequent 
case), or correlated with the airmass, the consequent flatfielding 
effects will have been removed by the linear or second order fit to 
the data mentioned previously.  Furthermore, data were rejected 
where brightness variations of CM Dra correlated with variations in the seeing.

In summary, care was taken to avoid error sources that could create
transit-like signatures in the data. However, such error sources could 
not be excluded with
certainty to be the source of any \emph{individual} feature in the
data that may appear interesting. Verification of transit features
is only possible by repeated observation of transits from the same
planet.

\begin{figure}
\centerline{\psfig{figure=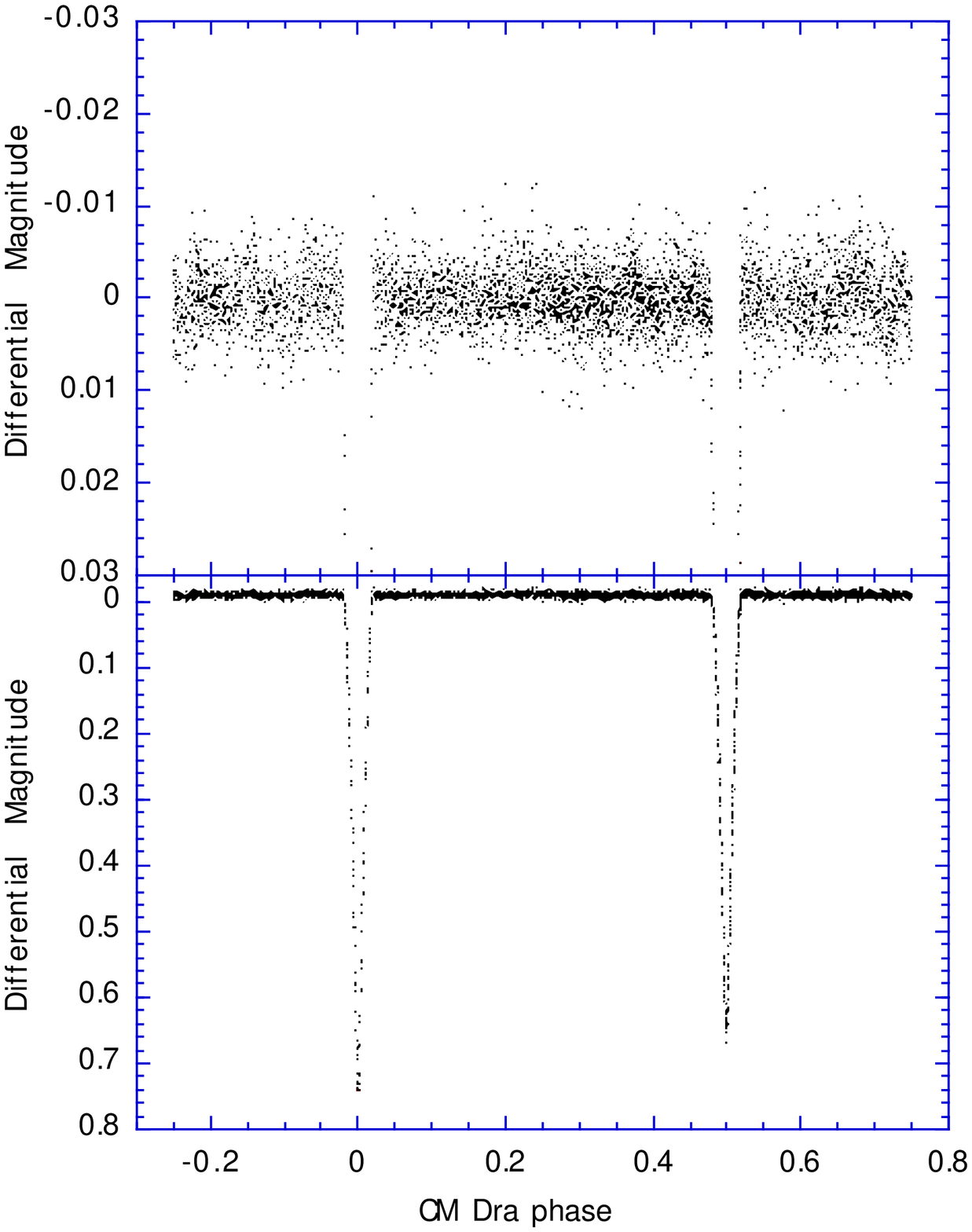,width=8.8cm}}
\caption{\label{fig:phase}Plot of the composite lightcurve of CM Dra 
against phase, containing the 17176 data points obtained in 1994-1996. The two panels 
show the same lightcurve with different magnitude scaling}
\end{figure}

\section{Results}
\begin{figure*}
%\vspace{20CM}
\centerline{\psfig{figure=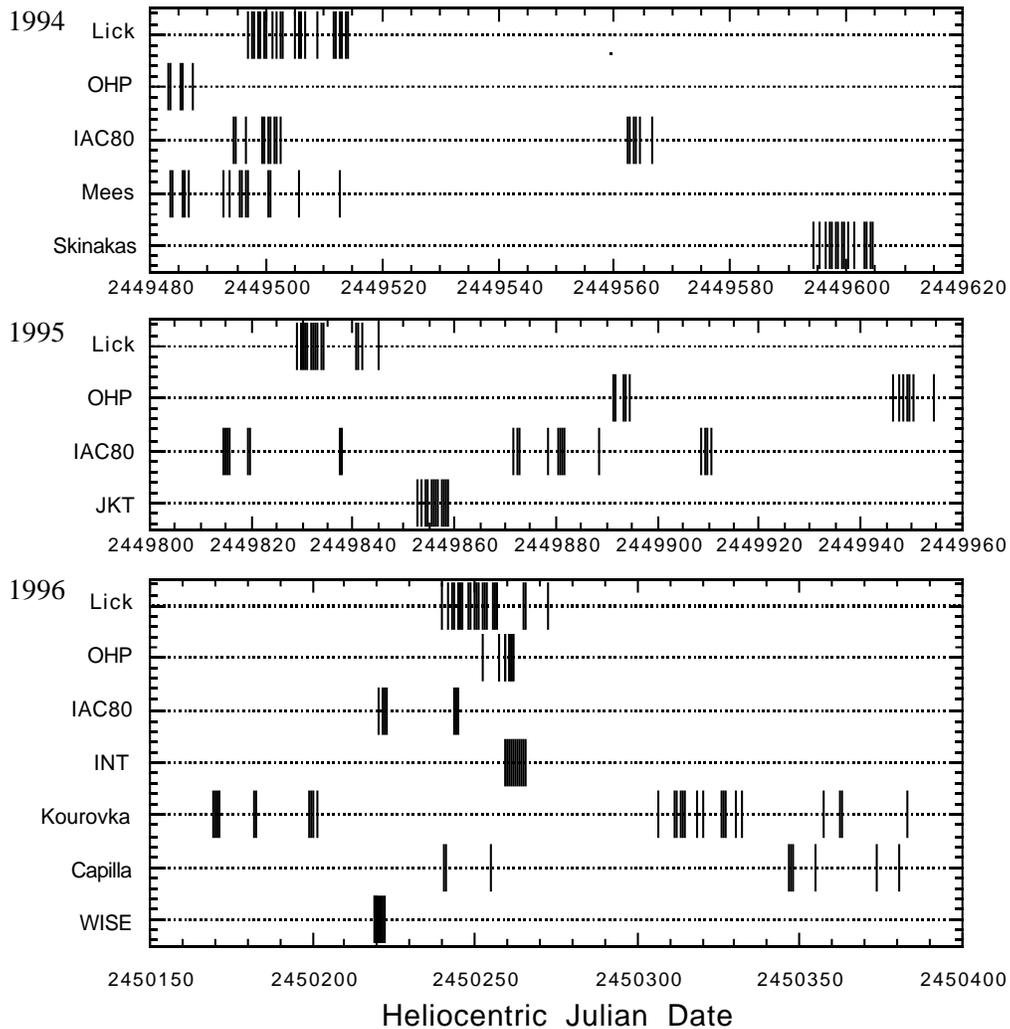,height=15cm}}
\caption{\label{fig:coverage}Coverage diagrams of the individual 
telescopes for 1994-1996}
\end{figure*}

The final lightcurve presented here contains 17176 points acquired
over three years, and gives a complete phase coverage for CM Dra
(Fig.~\ref{fig:phase}) at each of the three observational seasons. A
break-down of the observational coverage is given in
Table~\ref{tab:observations} and in Fig.~\ref{fig:coverage}.  The
'typical rms noise' column in Table~\ref{tab:observations} is the
noise of the final lightcurves from each telescope on good nights. On
some very good nights, the rms of the larger telescopes was better
than 2 mmag, whereas a noise of about 6 mmag was the cut-off for data
to be included into the final lightcurves.

\begin{table*}

\caption{\label{tab:observations} Overview of observational coverage 
for the 3 years}
\begin{center}
\begin{tabular}{lcccccccc} 
\hline
Telescope&\multicolumn{3}{c}{Observing coverage 
(hrs)$^1$}&&\multicolumn{3}{c}{number of data points}&typ. noise\\
\cline{2-4} \cline{6-8} 
&1994 &1995&1996&&1994&1995&1996&(mmag)\\
\hline
IAC 80cm&38&62&22&&1036&1101&473&4\\
JKT 1m&-&39&-&&-&765&-&3\\
INT 2.5m$^2$&-&-&42&&-&-&986&3\\
Mees 24''&53&-&-&&767&-&-&5\\
Capilla 24''&-&-&18&&-&-&298&4\\
Crossley 36''&65&50&46&&1670&2714&991&4\\
Kourovka 0.7m&&&68&&-&-&1926&5\\
WISE 40''$^3$&-&-&28&&-&-&390&4\\
Skinakas 1.3m&19&-&-&&1245&-&-&4\\
OHP 1.2m$^3$&11&34&22&&547&1452&815&3\\
%Korea 0.6m&-&7&-&&-&53&-&9\\
\hline
total &186&185&246&&5265&6032&5879&\\
\hline
\end{tabular}
\end{center}
$^1$Observing coverage is the length of the time for which a usable 
lightcurve of CM Dra was obtained. Coverage was considered continuous 
for interruptions of less than 15 minutes. $^2$Observations were 
taken through a redshifted H$\alpha$ filter. $^3$Some of these 
observations were taken in V band.
\end{table*}

\begin{figure}
\psfig{figure=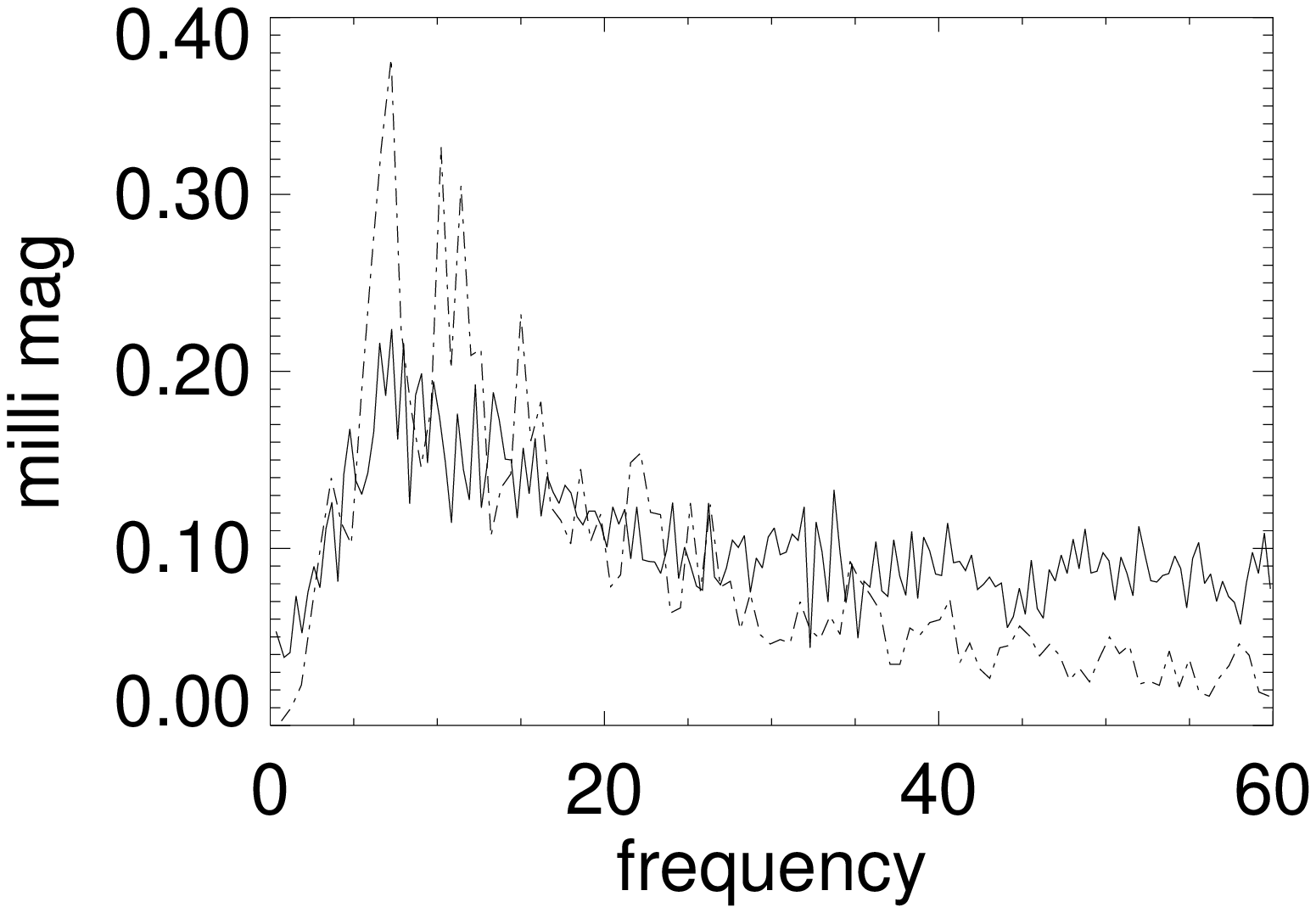,width=8.8cm}
\psfig{figure=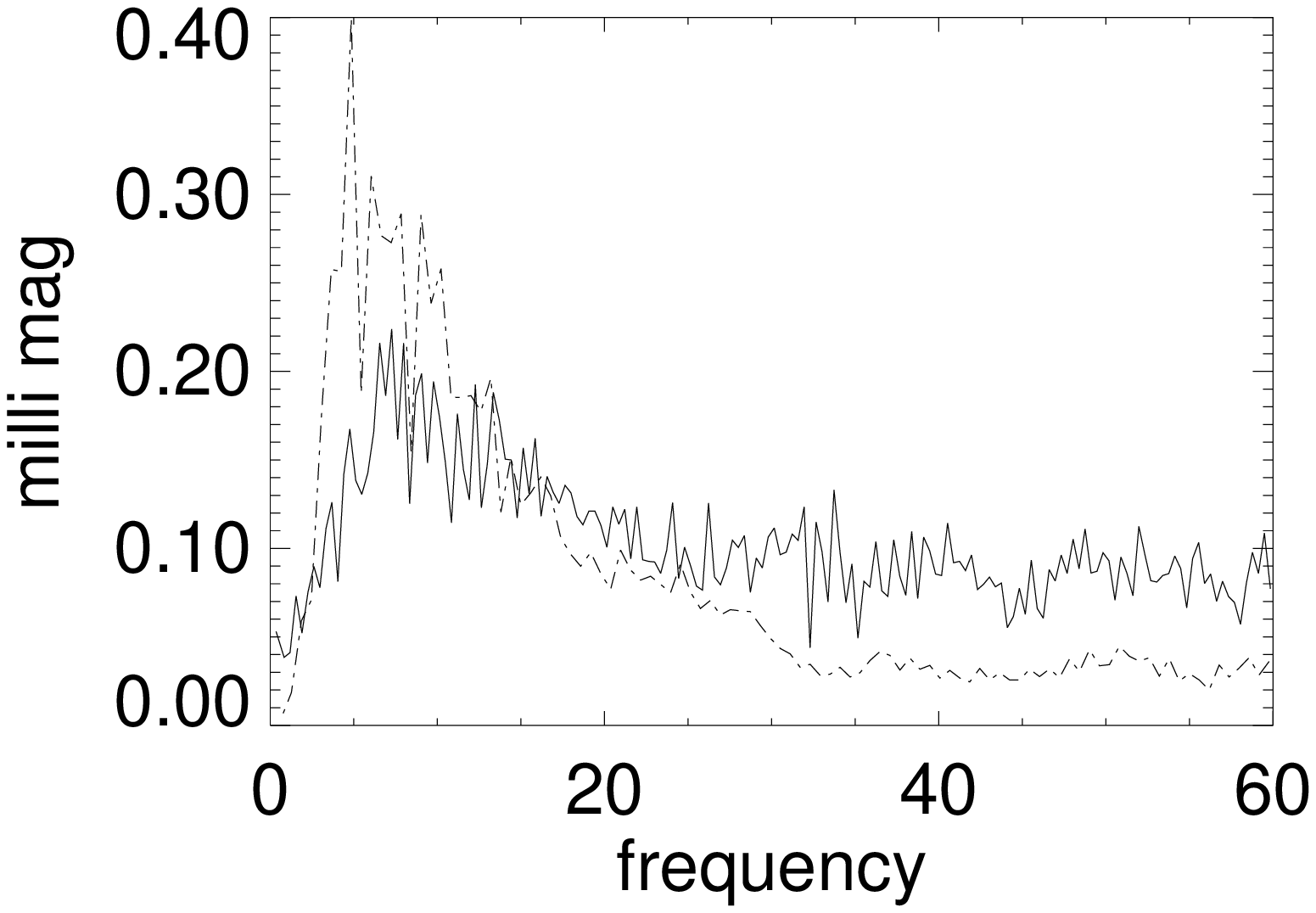,width=8.8cm}
\psfig{figure=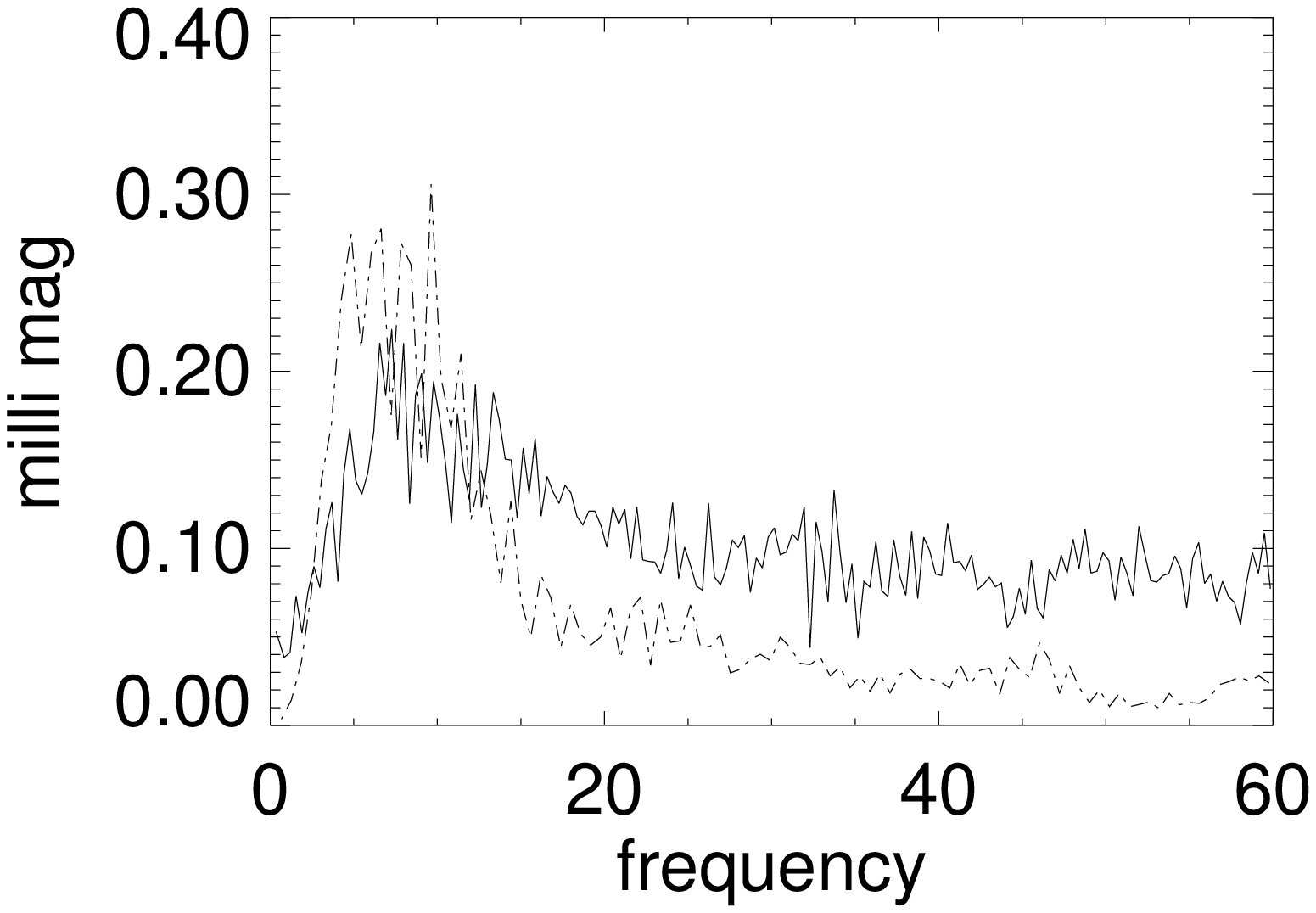,width=8.8cm}
\caption{\label{fig:lomb}Smoothed relative power spectra of the
observed lightcurve from 1994-1996 (solid line), over-plotted with
power spectra for models of planet transits with 10, 20 and 45 day
periods. The unit of the frequency axis is cycles per day. The
vertical scale is in milli-magnitudes, however it is important to note
that this scale applies only to the observed lightcurve. The power
spectrum of the model-transits cannot be scaled to measurable units,
since the duration of the transits is very short in comparison with
the time between transits, where the model lightcurve has no
signal. The models were calculated for the same time-points as the
observed data. For both the observed data and the model-lightcurve,
low frequent power (nightly extinction) was removed in the same way
(see description in Sect. 3). In all cases, the power of the modeled
planetary transits is concentrated towards frequencies of 5-15
day$^{-1}$, whereas the power from the noise in the observed data is
relatively flat.}
\end{figure}

\subsection{The noise of the lightcurve}
Figure~\ref{fig:lomb} shows power spectra of the observed data
over-plotted with spectra from model planetary transits with periods
of 10, 20 and 45 days. The power spectra were calculated by phase
dispersion minimization with the program POWER (Kjeldsen, personal
communication). For the detection of planetary transits which may be
hidden from ordinary view in the noise of the lightcurve, it is
important to note the differences in the power spectra. The
observational data have a relatively flat spectrum, with a maximum at
frequencies around 7 $day^{-1}$, corresponding to a period of 200
minutes. The power of the planetary transits peaks much more
pronounced between 5 and 12 $day^{-1}$ (corresponding to periods
between 1 and 2.5 hours, which is the typical length of planetary
transits), and there is little power left at frequencies above 20
$day^{-1}$. For this analysis, the mutual eclipses of CM Dra, as well
as the nightly extinction slopes have been removed which accounts for
the absence of power at frequencies below 5 $day^{-1}$. This also
fortuitoisly removes most of the power from signals that appear with
the period of CM Dra (such as starspots). As can be seen in
Fig.~\ref{fig:lomb}, there is very little power left at CM Dra's
period of 0.79 $day^{-1}$. Also absent is the first harmonic at 1.58
$day^{-1}$, which might be strong, since primary and secondary eclipses
have nearly equal amplitudes.

\begin{table*}
\caption{\label{tab:flares} List of flares observed in 1994-1996}
\begin{center}
\begin{tabular}{ccccccc} 
\hline
Julian Date&duration&amplitude&nightly rms&telescope&remarks\\
(jd-2400000) & (days)&(mag)&(mag)& & \\
\hline
49483.587&$>$0.021&0.023&0.0075&OHP&end of flare not observed \\
49495.78&0.036&0.017&0.0050&Mees& \\
49563.475&0.030&0.027&0.0042&IAC80&first peak of double flare\\
49563.505&0.025&0.026&0.0042&IAC80&second peak of double flare\\
49598.396&0.035&0.018&0.0035&Skinakas&clear flare in good night\\
49600.379&0.021&0.030&0.0044&Skinakas&clear flare in good night\\
49601.409&0.025&0.05&0.0041&Skinakas&clear flare in good night\\
49854.495&0.025&0.046&0.0026&JKT&clear flare in good night\\
49891.440&0.020&0.041&0.0027&OHP&clear flare in good night\\
50238.848&0.035&0.020&0.0030&Lick&uncertain, as shortly after clouds\\
50243.757&0.025&0.020&0.0026&Lick&first of two flares that night\\
50243.872&0.017&0.013&0.0026&Lick&uncertain second flare, only 1 high 
point\\
50244.510&0.030&0.018&0.0058&IAC80&noisy night\\
50262.615&0.009&0.025&0.0035&INT&unclear, at begin of secondary 
eclipse\\
50263.464&0.026&0.047&0.0036&INT&first peak of double flare\\
50263.502&0.018&0.035&0.0036&INT&first peak of double flare\\
\hline
\end{tabular}
\end{center}
%Notes to Table
\end{table*}

\subsection{Eclipse minima timing}
An analysis of the minimum times of the mutual eclipses of an
eclipsing binary may reveal the presence of a third body in this
system.  In such a case, the orbital period of the third body should
cause periodic changes in the time of the minima, as the distance to
the binary system is offset by its motion around the 3-body
barycenter. In the CM Dra system, for example, a planet with the mass
of Jupiter at a distance of 5 AU would cause a periodic shift of
minimum times with an amplitude of 5.5 seconds (\cite{doyl97}).  This
method is however unsuitable for the detection of planets with masses
significantly smaller than giant planets. The three years of
observational coverage of CM Dra contain 16 primary and 19 secondary
eclipses, on which the time of minimum brightness could be measured
reliably, with uncertainties of less than 10 seconds. Minimum times
were measured with the 7 segment Kwee-Van Woerden method
(\cite{kvw}). We also re-measured the minimum times of the three
eclipses observed by Lacy (1977).  Against the epochs cited by Lacy,
re-measuring gave discrepancies of 10 and 35 seconds for his two
primary eclipses, and a discrepancy of 6 seconds against the one
secondary he observed.  We therefore prefer to assign new epochs to CM
Dra as follows: primary eclipse: JD 2449830.757\,00$\pm$0.000\,01,
secondary eclipse: JD 2449831.390\,03$\pm$0.000\,01, and a period of
1.268\,389\,861$\pm$0.000\,000\,005 days, based on a fit to the 35
minimum times from 1994 to 1996 and the remeasured values for Lacy's
primary eclipses. With these new elements, our observed~-~calculated
(O-C) minimum times have a scatter of only about 6 seconds. This small
scatter excludes periodic changes in the minimum times with amplitudes
larger than 9 seconds and periodicities of less than about 4 years. We
therefore cannot support the claim made by \cite{guin98} of a
70.3 day periodic variation in minimum times with an amplitude of 18
seconds, that would have been indicative of a very massive planet (see
also \cite{deeg98}).

%An O-C diagram based on these
%elements is shown in Fig.~\ref{fig:OCdiag}. No periodicity can be
%derived from this diagram, of note however is an increase in O-C of
%about 25 seconds throughout the 130 days (105 orbits) being spanned 
%by observations in 1996.
 
%\begin{figure*}
%\vspace{10CM}
%\caption{\label{fig:OCdiag} O-C diagram of the observations from 
%1994-1996, based on the elements given in the text. The abscissa 
%gives the number of orbits of CM Dra relative to the reference 
%epoch, 
%the ordinate the time difference in seconds. Primary eclipses are 
%indicated as squares and secondary eclipses as diamonds. The error 
%bars are those given by the 7 segment Kwee V. Woerden method, they 
%do 
%{\it not} include potential systematic errors at some telescope 
%acquisition systems.}
%\end{figure*}

\subsection{Flares}
Several flare events of CM Dra were observed. In Table~\ref{tab:flares}, the time of the peak maximum,
the total duration, maximum brightness, the average noise(rms) of 
that night,
and the observing telescope are given. Lacy (1977) already noted the
low flare rate of CM Dra compared against the rate of $2 hr^{-1}$
expected for a Population I flare star. Our observations confirm a low flare
rate of $\approx 0.025 hr^{-1}$ over the 3 years of observations,
although only 2 flares were observed in 1995, corresponding to a rate
of $\approx 0.011 hr^{-1}$. Since only flares with durations of more
than several minutes (identified flares had to contain more than
one data-point, as spikes of single points may also be caused by
cosmic rays hitting near CM Dra) and with amplitudes of $
\stackrel{>}{_\sim} 0.015 mag$ could be identified clearly, the
observed flares represent a minimum flare rate. Since CM Dra is
a tidally locked system, we cannot derive any information about its
age from the flare rate. However, for a relatively fast rotator, CM
Dra's flare rate is notably low.

Since flares introduce a spurious signal into the lightcurve, they
have been removed from the final lightcurves that are being used in
the further analysis to detect planetary transits.

%clear page to force printing of figs
%\clearpage

\begin{figure*}
%below, for preprint (2 col) style
\centerline{\psfig{figure=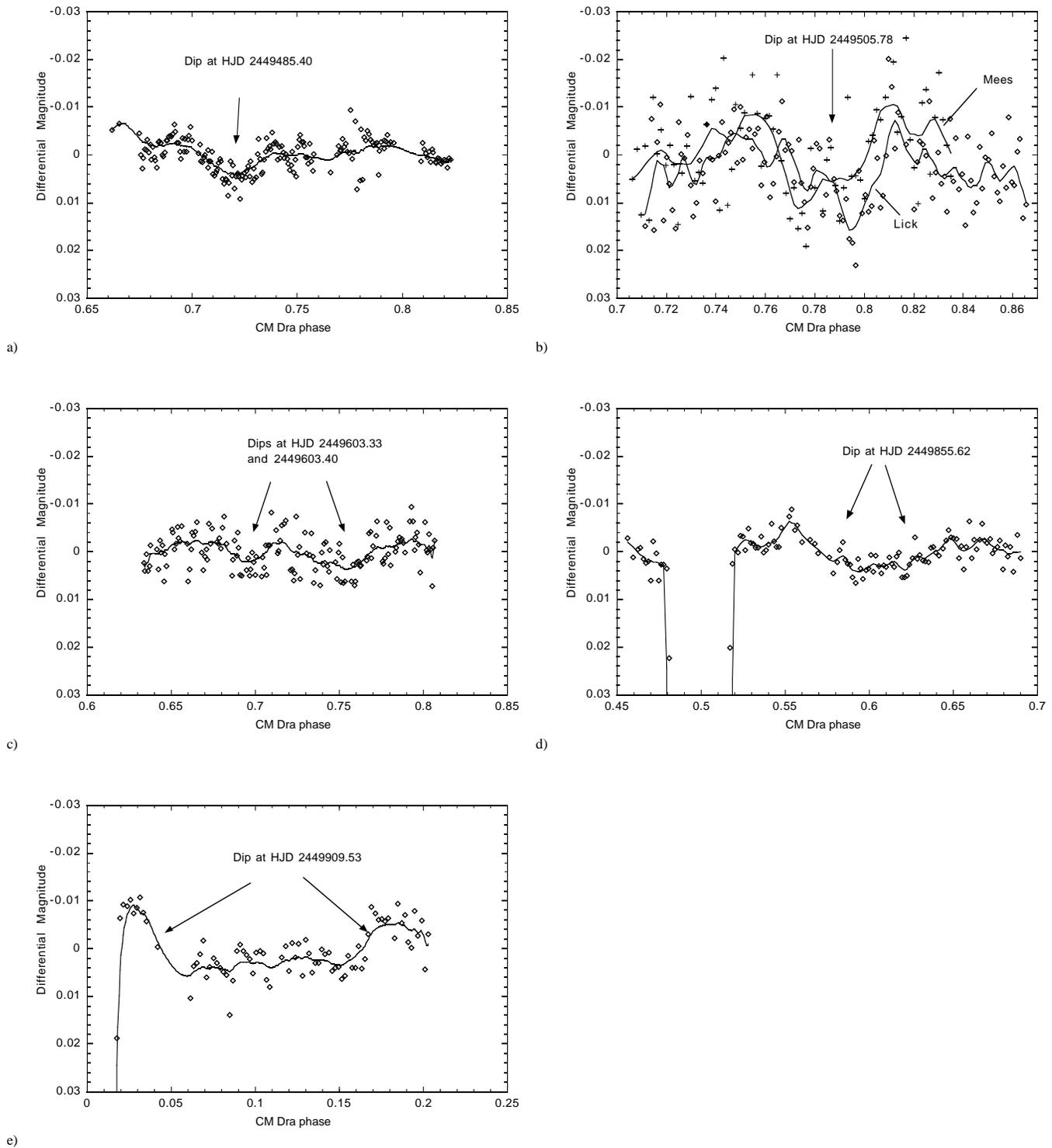,width=18cm}}
%below, for referee (1 col) style
%\centerline{\psfig{figure=H0739F6.eps,width=15cm}}
\caption{\label{fig:dipevents} The six planetary transit event
candidates from Table~\ref{tab:diplist}. The lightcurves are plotted
against the phase of CM Dra. The data are shown as squares; the line
indicates a smoothing fit to the data. a) Event at Heliocentric JD
2449485.395 observed at OHP. b) The event centered at JD 2449505.78,
observed simultaneously at Lick (squares) and Mees (crosses)
observatories. The light drop appears about 10 minutes later in the
Lick observations; this delay may be caused by noise in the data. The
amplitude in the Lick data is 0.001 mag, whereas in the Mees data it
is 0.0014 mag. c) Double dip at JD 2449603.33 and 2449603.40. d) Dip
at JD 2449855.62, occurring shortly after a secondary eclipse. e) A
long flat dip observed at JD 2449909.53. This event occurred shortly
after a primary eclipse that was observed at the beginning of this
night, and had a duration of 180-240 minutes.}
\end{figure*}

\subsection{Potential planetary transits} 
%STRONGLY MODIFIED ON 23 JAN	
The goal of these observations has been the detection of transits from 
planets orbiting the CM Dra system.  The lightcurves were therefore 
visually scanned for the presence of events which might be indicative 
of planetary transits.  Such events, further called \emph{transit 
candidates}, are typified by being temporary faintenings of CM Dra's 
brightness by a few millimagnitudes, with normal durations of 45 - 90 
mins.  Transit events may last as long as a few hours, but only if CM 
Dra is very close to a primary or secondary eclipse in the middle of 
the event.  Several potential transit events have been observed and 
are included in Table~\ref{tab:diplist}.  Their light-curves are shown 
in Fig.~\ref{fig:dipevents}.  It is not possible to derive elements of 
potential planets (with the exception of their diameter), if any 
\emph{one} of these transit candidates is considered isolated, since 
the duration of a planetary transit depends only weakly on the 
planetary period.  An exception is the transit candidate at JD 
2449909.53 (Fig.~\ref{fig:dipevents}e), where the long transit duration 
is only compatible with a planet with a period of less than 9 days, or 
a period between 25 and 32 days.  Even if a hypothetical planet has 
already caused two observed transits, its period will generally still 
be ambiguous because of the unknown number of orbits completed between the two 
transits.

\begin{table*}
\caption{\label{tab:diplist} List of photometric events which might 
be caused by 
planetary transits}
\begin{center}
\begin{tabular}{cccccc} 
\hline
Julian Date&duration&amplitude&nightly rms&telescope&remarks\\
(jd-2400000) & (days)&(mag)&(mag)& & \\
\hline
49485.395&0.046&0.007&0.0065&OHP&\\
49505.775&0.05&0.01&0.0077&Mees&observed also at Lick?\\
49505.785&0.05&0.008&0.0065&Lick&observed also at Mees?\\
49603.33&0.033&0.004&0.0038&Skinakas&first of double dip?\\
49603.399&0.06&0.005&0.0038&Skinakas&second of double dip?\\
49855.620&0.098&0.005&0.0022&JKT&\\
49909.535&0.16&0.01&0.0032&IAC80&long flat dip\\
\hline
\end{tabular}
\end{center}
%Notes to Table
\end{table*}

\begin{figure}
%\vspace{1CM}
a)\psfig{figure=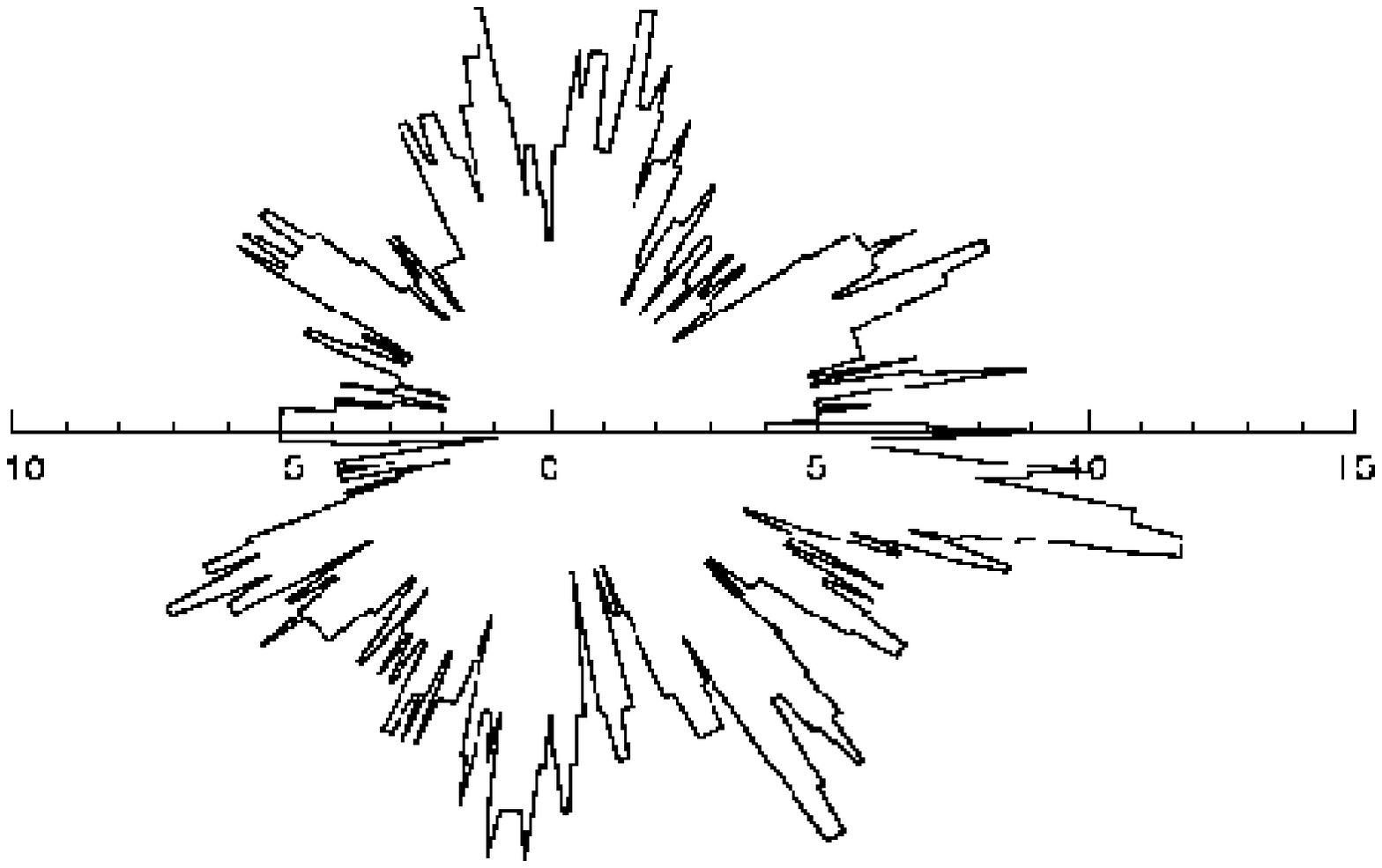,width=8.8cm,height=8.8cm,rheight=7.5cm}\linebreak
b)\psfig{figure=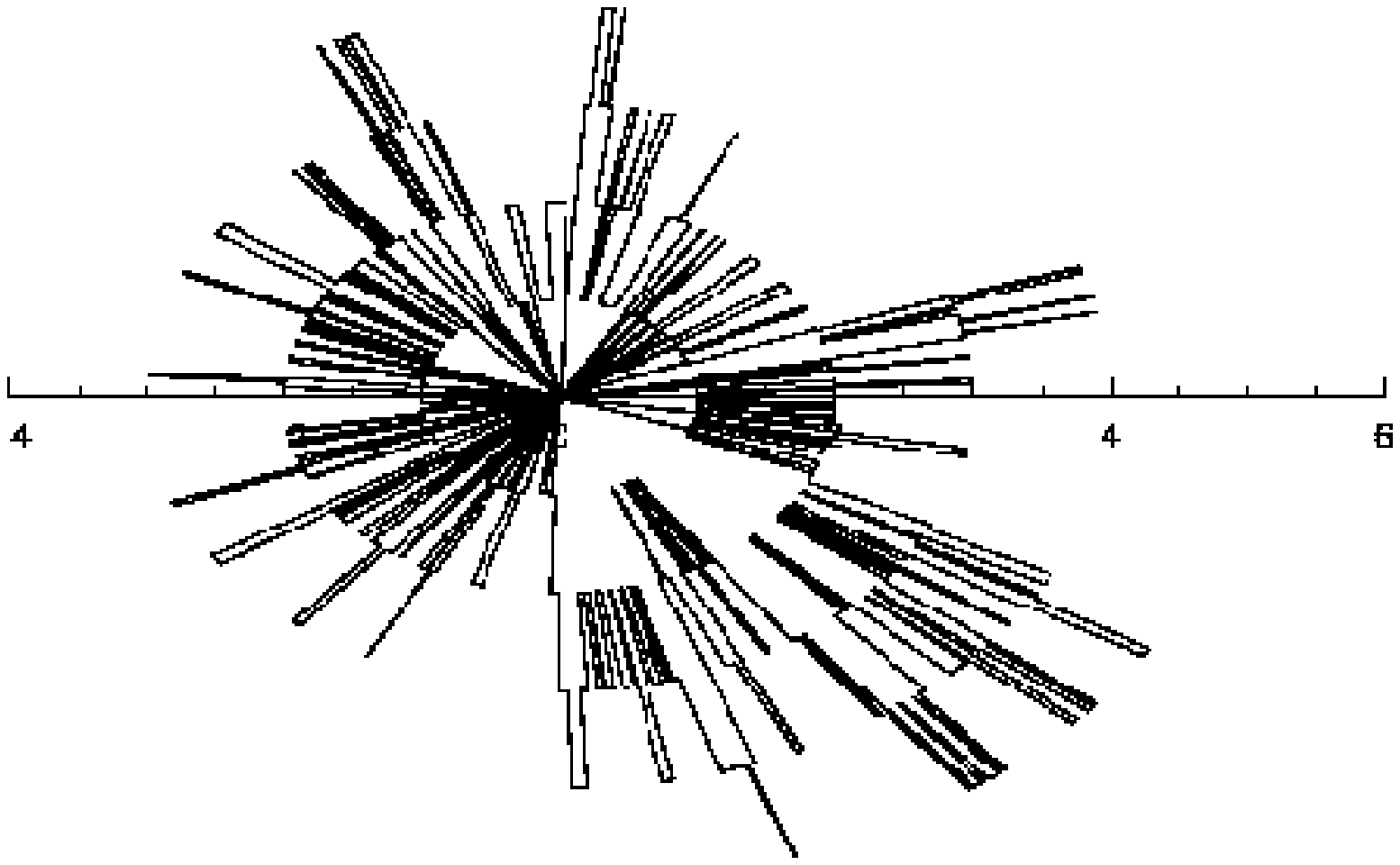,width=8.8cm,height=8.8cm}
\caption{\label{fig:transitcoverage}
The expected number of transits that would have been observed in our 
actual lightcurve of 617 hrs, for a hypothetical planet with a period 
of a) 10.14 days and b) 45.14 days. The clockwise direction is the 
planet's phase (going from 0 to 360 degrees) and the radial direction gives 
the number of transits. The phase is defined here as the phase of the 
planet at JD 2450000.0; phase zero means the planet is in front of 
the barycenter of CM Dra).}
\end{figure}

{\it If} there are any short-period ($\stackrel{<}{_\sim}$ 60 days)
planets around CM Dra, it would not be unlikely that these have
already been observed more than twice in our 617 hrs of observational
coverage. For example, a planet with a period of 10 days will cause 2
transits (one transit for each component of CM Dra) every 240 hrs,
giving an average of 5.1 observed transits within the 617 hours. The
exact number of observed transits depends, of course, on the exact
period and orbital phase -or epoch- of a planet (at 0\degr phase, the
planet is crossing in front of the binary barycenter). The numbers of
transits that are expected in our actual lightcurve are given in
Fig.~\ref{fig:transitcoverage} for two examples of hypothetical
planets with periods of 10.14 and 45.14 days (the odd periods were
chosen to avoid aliasing effects). For a 10.14 day planet, the
observed lightcurve would contain between 1 and 12 transits, depending
on the planet's phase. A 45.14 day planet could have caused up to 5
transits, but there is also a 15\% chance of missing this planet
entirely. The probabilities that certain numbers of transits from
planets with various periods are within the 617 hrs observational
coverage is given in Table~\ref{tab:transitprob}.

\begin{table}
\caption{\label{tab:transitprob} The probabilities (in percent) of 
the number of transits observed within the observational coverage 
from planets with selected periods. For example, if there is a planet 
with a period of 30.14 days present, the probability that it has 
caused 1-2 transits in the observations from 1994-1996 is 53 \%}
\begin{center}
\begin{tabular}{cccc} 
\hline
Planet&No transits&1-2 transits&more than 3 transits\\
period&(\%)&(\%)&(\%)\\
\hline
10.14 days&0.0&2.4&98\\
15.14 days&2.1&13&85\\
20.14 days&2.4&31&67\\
30.14 days&5.7&53&41\\
45.14 days&15&63&22\\
60.14 days&20&69&10\\
\hline
\end{tabular}
\end{center}
%Notes to Table
\end{table}

\begin{figure}
\centerline{\psfig{figure=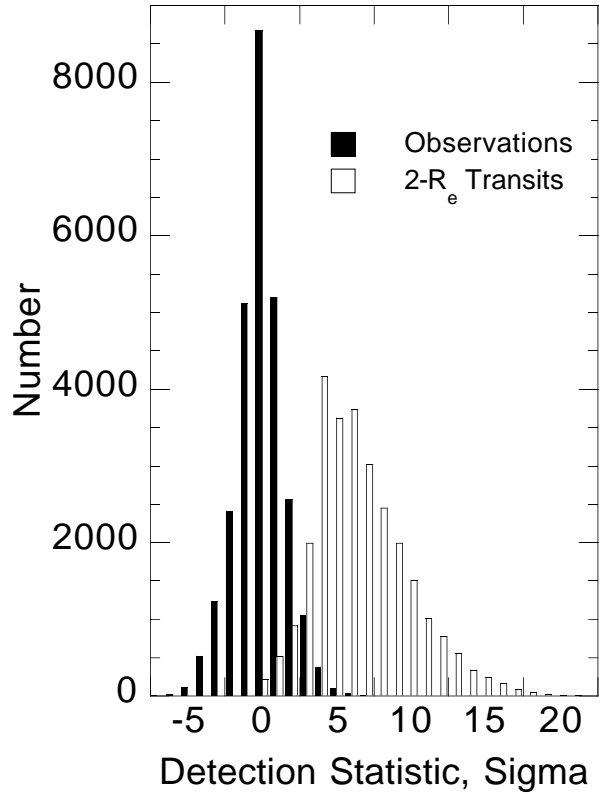,width=8.8cm}}
\caption{\label{fig:detcstat} 
Shown are the generated detection statistics for a hypothetical 
2-Earth-Radii planet (right) that would cause transits of 1.5 hours each 
in our observational light curve (left). The degree of 
overlap of the two peaks (detection on the right and non-detection on 
the left) is a measure of detectability of the transit signals - if they
overlap completely, no detection is possible. Clearly a detection can
take place reliably at least 50\% of the time, demonstrating that we have
already reached a detection limit well into the terrestrial-sized planet 
range (a 2-Earth-Radii Planet is 1\% the size of Jupiter).
}
\end{figure}

%BELOW INCLUDED ON 23 JAN
Assuming that we may have already observed transits of a potential
planet 3 or more times, we searched for periodicities among the
transit candidates, using those shown in Fig.~\ref{fig:dipevents}, and
a few less pronounced candidates.  Among these \emph{transit}
candidates, several thousand possibilities for \emph{planetary}
candidates (with 3 or more already observed transits!)  were found.
Each \emph{planetary} candidate represents a combination of a possible
planetary orbital period and epoch, the later one being defined as the
time, when the planetary candidate is crossing in front of the
barycenter of CM Dra, as stated.  Modeled lightcurves of these
planetary candidates (assuming radii of 1.5, 2 and 2.5 R$_{E}$, and
using the model-code that generated the curves shown in
Fig.~\ref{fig:transshapes}) were cross-correlated against the observed
lightcurve.  This led to a list of several 100 planetary candidates
which reasonably fitted the observed lightcurve.  For the best 10 of
these planetary candidates, transit times were predicted, and pointed
observations at these predicted times were undertaken in the Spring of
1997 at the IAC80 telescope.  Unfortunately, no transits were observed
at any of these predicted times. Except for these 10 tested planetary
candidates, however, these results cannot rule out any other planets
in this size range yet. Since it is impossible to observe at predicted
transit times for the whole list of several 100 planetary candidates,
we are currently again engaging in observations that will increase
transit coverage in general, in order to reveal further smaller
transit candidates.  We want to emphasize, that the observed transit
candidates of Fig.~\ref{fig:dipevents} are examples of events that can
be caused by planetary transits, but only \emph{repeated} observation
of transits from the \emph{same} planetary candidate can verify their
true nature.

Whereas there are multitudes of planetary candidates that may have
caused the light drops reported in Table~\ref{tab:diplist}, we note
that there are no observed light-drops with amplitudes larger than
0.01 mag.  Cross-correlations between model-lightcurves of planetary
transits and the observed light-drops showed that in no case can
planets much larger than 2.5 R$_{E}$ be responsible for these observed
light-drops. Our observational coverage gives a confidence of about
80\% that such larger planets with periods of less than 60 days can be
excluded. For periods of less than 20 days, this confidence is 98\%.
If the light-drops of Table~\ref{tab:diplist} and
Fig.~\ref{fig:dipevents} are indeed from planetary transits, they must
result from planets with sizes between 1.5 and 2.5 R$_{E}$.

A signal detection approach was taken for preliminarily assessing
confidence in the detectability of planets in this size range within
the current data set. For this, 36000 model transits were generated
for 2 R$_{E}$ planets with periods of 10 days through 30 days (the
habitable zone around CM Dra), and included in the data.  Subsequent
detection attempts (Fig.~\ref{fig:detcstat}) showed that 50\% of the
time, the detection statistic for a 2 R$_E$ planet transiting CM
Dra would be above the detection statistics generated from our
photometric data (the number of times the cross-correlation values for
the transit curve were higher than the noisy observational data).
Thus, our data from the 1994 through 1996 observing seasons allow us a
confidence level of 50\% that an actually existing 2 R$_{E}$ planet
would have been detected. For a 3 R$_{E}$ planet, a similar test gave 90\%
confidence.

We would like to note, that one transit candidate reported by our 
group (\cite{iauc6425}) turned out to result from a problem with the 
flatfielding of that night's images.  There were also reports 
(\cite{guin96}, 1997) of CM Dra being fainter by 0.08 mag throughout 
the whole night of June 1, 1996.  If this event would have been caused 
by a planet with about 0.94 Jupiter diameters, the long duration of 
the transit would indicate a planet with an orbital period of about 
2.2 years.  For such long periodic planets, the probability that their 
orbital plane crosses in front of CM Dra, as well as the probability 
of catching a transit at the right time, are both very low.  Most 
important however, such a planet should have caused periodic 
variations in the minimum times of CM Dra's primary eclipse of over 10 
seconds, which have \emph{not} been detected (see Sect. 4.2).  We 
never encountered any brightness variations this large and this long 
in duration (the latter would have been due to our reduction 
procedure, where we set the brightness of CM Dra relative to its 
reference stars to zero for each night).  Our data are therefore
less sensitive to slow-changing atmospheric extinction variations but 
would miss unusually long transits that began before \emph{and} ended 
after the nightly observations.  However, since none of the reference 
stars anywhere near the field has a red color similar to CM Dra, 
relative brightness changes may have been caused by differential color 
extinction.  In addition, Guinan et al.'s observations were done in I 
band, which is notorious for variations caused by OH$^{-}$ in the 
atmosphere.  We believe therefore, that the event reported by them was 
most likely caused by a night with abnormal extinction, or was due to 
a flatfielding problem.

\section{Summary}

The TEP network has embarked on obtaining extended observational coverage of
the eclipsing M star binary CM Dra with the goal of detecting transits
of planets which are orbiting in the plane of the binary components.
Reported here are the data from the first three years of this project,
covering the years 1994 through 1996. Time series of about 200 hours
coverage were obtained in each year of observations, containing a
total of 17176 data points. This gives the most thorough coverage of
any binary eclipsing star observed to date. In this coverage we
observed six events in which the brightness of CM Dra dropped
between 0.004 mag and 0.01 mag over a time scale of an hour. These
events are compatible with planetary transits, but their true nature
can only be ascertained by further observational coverage. The absence
of any light drops larger than 0.01 mag allows us to rule out the
presence of close planets with radii of larger than 2.5  R$_{E}$ 
(corresponding to about 1.1\% the volume of Jupiter), with a
confidence of 80\% for orbital periods of less than 60 days. The
confidence is 98\% that such planets are absent with periods of less
than 20 days.  The light-loss events listed in 
Table~\ref{tab:diplist} are examples of what lightcurves from the 
transits of extrasolar planets with sizes between 1.5 and 2.5 
R$_{E}$ should look like. But again, their true nature can only be 
confirmed by observations of repeated transits from the same planet.

Furthermore, the light-loss events in Table~\ref{tab:diplist} only
contain events that are {\it visible} in the lightcurve. The typical
noise in the lightcurve is 2-3 mmag, and events from planets smaller
than about 2 R$_E$ could be hidden in the lightcurve. Such small
planets, although their transits are visually undetectable in the
lightcurve, may be found by the application of a matched filter
algorithm - i.e. cross-correlations with the quasi periodic signals
caused by planetary transits (\cite{jenk96}). Detection statistical
calculations show, that the confidence that planets with 2R$_E$ could
be detected in the current data - \emph{if} they are present - is
about 50\%.  In conclusion, it is therefore definitively possible to
use the photometric transit method with ground based 1~meter class
telescopes to obtain results about the presence of extrasolar
terrestrial-sized planets within a star's habitable zone. With larger
telescopes and longer observing times, this method may be extended to
the detection of terrestrial-like planets around many other eclipsing
binary systems, as well.

\acknowledgements The TEP network wishes to thank numerous people, 
without whose support the acquisition of these data would have been 
impossible.  The observations by LRD and RS at the oldest reflecting 
telescope still in professional use - the Crossley Telescope at Lick 
Observatory - were very support intensive and we thank Ellen Blue, 
Moira Doyle, Neil Heather, Dave Koch, and the time allocation 
committee at UC Santa Cruz for their support.  For observations and 
reductions at Capilla Peak, we thank Randy Grashuis.  We thank Ayvur 
Akalin who helped us with an implementation of the Kwee-v.  Woerden 
algorithm.  We also thank Woo-Baik Lee and Ho-Il Kim at the Korean 
Astronomy Observatory for helpful correspondence. 

The IAC80 telescope is operated at Iza\~na Observatory, Tenerife by the 
Instituto de Astrofisica de Canarias.  We thank the University of 
Rochester for use of the Mees Telescope.  HJD acknowledges a 
postdoctoral fellowship from the Rochester Institute of Technology in 
1994, and later a grant 'Formaci\'{o}n de Personal Investigador' from 
the Ministry for Education and Culture of Spain.  The observations by 
LRD were supported in part by a grant from the NASA Ames Research 
Center Director's Discretionary Fund.  JS acknowledges funding 
provided by the French 'Programme National de Planetologie (CNRS)'.

\clearpage

\end{document}